\documentclass[global]{svjour}

\usepackage{latexsym}
\usepackage{graphics}

\newcommand{\He}{$^4$He$^*$}

\journalname{}
 
\begin{document}

\title{Simple method for producing Bose-Einstein condensates of metastable helium using a single beam optical dipole trap}
\titlerunning{Simple method for producing BEC of \He~using a single beam ODT}

\author{Adonis Silva Flores \and Hari Prasad Mishra \and Wim Vassen \and Steven Knoop}
\institute{LaserLaB, Department of Physics and Astronomy, Vrije Universiteit Amsterdam, De Boelelaan 1081, 1081 HV Amsterdam, The Netherlands}
\date{\today}

\maketitle

\begin{abstract}
We demonstrate a simple scheme to reach Bose-Einstein condensation (BEC) of metastable triplet helium atoms using a single beam optical dipole trap with moderate power of less than 3~W. Our scheme is based on RF-induced evaporative cooling in a quadrupole magnetic trap and transfer to a single beam optical dipole trap that is located below the magnetic trap center. We transfer $1\times 10^6$ atoms into the optical dipole trap, with an initial temperature of 14~$\mu$K, and observe efficient forced evaporative cooling both in a hybrid trap, in which the quadrupole magnetic trap operates just below the levitation gradient, and in the pure optical dipole trap, reaching the onset of BEC with $2\times 10^5$ atoms and a pure BEC of $5\times 10^4$ atoms. Our work shows that a single beam hybrid trap can be applied for a light atom, for which evaporative cooling in the quadrupole magnetic trap is strongly limited by Majorana spin-flips, and the very small levitation gradient limits the axial confinement in the hybrid trap. 
\end{abstract}

\section{Introduction}\label{Introduction}

Quantum-degenerate atomic gases in optical dipole traps provide the starting point of many experiments. To realize these samples one can directly load a laser-cooled sample into an optical dipole trap (ODT) and perform evaporative cooling, which however requires very high ODT powers to provide sufficient trap volume, depth and confinement. Alternatively, one loads the atoms first in a magnetic trap and performs evaporative cooling, and afterwards transfers a dense and compact atomic cloud into an ODT, which now requires a much lower ODT power, at the expense of experimental complexity. Within this last category a very elegant approach is the hybrid trap (HT), introduced in Ref.~\cite{lin2009rpo}, consisting of a simple quadrupole magnetic trap (QMT) and a single beam ODT. Efficient evaporation and Bose-Einstein condensation (BEC) have been demonstrated (see e.\,g.\,\cite{lin2009rpo,kleinebuning2010asg,mishra2015epo}), using ODT powers of only a few Watts. By switching off the QMT completely the atoms are transferred from the HT to a pure ODT. 

The hybrid trap has been mostly applied to $^{87}$Rb, but is assumed to be generally applicable to other magnetically trappable atomic species \cite{lin2009rpo}. However, the application of HT strongly depends on the mass of atom. Most importantly, the rates of Majorana loss and heating, which determine the temperature that can be reached by evaporative cooling in a QMT, scale inversely with mass \cite{petrich1995stc,dubessy2012rbe}. This limits the transfer efficiency for light atoms, or puts constraints on the trap volume and trap depth, and therefore the power, of the ODT. Furthermore, for light atoms evaporative cooling in the HT is limited as the additional axial confinement provided by the QMT is small because of the small levitation gradient, below which the QMT has to operate in the HT. Finally, the small levitation gradient puts experimental limits on the control of the displacement of the QMT with respect to the ODT, which further limits the axial confinement.
 
Here we report on the production of a metastable triplet helium (\He) BEC using a single beam HT with a moderate power of less than 3~W, demonstrating the application of HT for a light atom. Our work provides a novel and simple method for obtaining a \He~BEC, which can be used for atom optics experiments \cite{schellekens2005hbt,jeltes2007cot,hodgman2011dmo,lopes2015aho,manning2015wdc} or precision spectroscopy for fundamental tests of two-electron quantum electrodynamic theory \cite{rooij2011fmi,notermans2014hps,henson2015pmo}. So far, \He~BECs have been obtained in Ioffe-Pritchard or cloverleaf type of magnetic traps \cite{robert2001bec,pereira2001bec,pereira2002poa,tychkov2006mtb,dall2007bec,doret2009bgc,keller2014bec}, which has been subsequently transferred to a single beam \cite{partridge2010bec,dall2010tmi} or crossed beam ODT \cite{rooij2011fmi} (see Ref.~\cite{vassen2012cat} for a review on experimental work on ultracold \He). Very recently a \He~BEC has been realized in a crossed beam ODT \cite{bouton2015fpo}, using a total power of 26~W, in which the ODT is loaded from a QMT, following evaporative cooling to BEC in the ODT. 

This paper is organized as follows. In Sect.~\ref{HT} we give a brief description of HT for the particular case of metastable helium. In Sect.~\ref{expsetup} we describe our experimental setup and initial cooling scheme to load the single beam HT or ODT. In Sect.~\ref{results} we present our results regarding the alignment and loading of the HT, comparing evaporative cooling in the HT and ODT, and provide evidence for BEC. Finally, in Sect.~\ref{conclusions} we conclude. 

\section{Single beam hybrid trap for \He}\label{HT}

In the HT a single beam ODT is aligned slightly away from the QMT center, such that the trap minimum of the combined magnetic and optical trapping potential is at a finite magnetic field, and atoms do not suffer Majorana spin-flip losses once loaded in the HT. This also means that the atoms remain spin-polarized, which is crucial for \He~in order to avoid strong losses due to Penning ionization. After forced evaporative cooling in the QMT the magnetic field gradient of the QMT is ramped down such that the vertical gradient is equal or lower than the levitation gradient $B'_{\rm lev}\equiv m g / \mu$, at which the QMT alone cannot trap atoms. Here $m$ is the mass, $g$ is the gravitational acceleration and $\mu$ is the magnetic moment of the atom. Lowering the power in the ODT beam allows further evaporative cooling in the HT, in which the ``hot atoms'' can escape mainly downwards. 

\begin{figure}
\center
\resizebox{0.95\textwidth}{!}{%
	\includegraphics{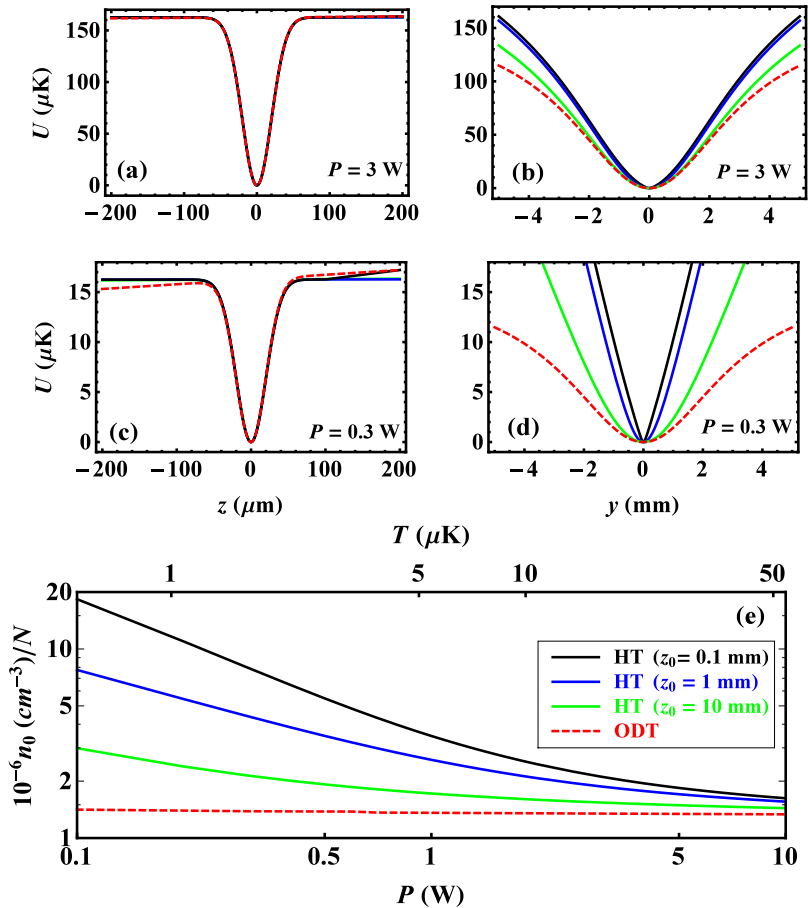}
}
\caption{(Color online) Trapping potentials and peak densities of HT (solid lines) and ODT (red dashed lines) for \He, calculated for $B'=B_{\rm lev}=0.35$~G/cm, $\lambda=1557$~nm and $w_0=40$~$\mu$m (corresponding to $y_R=3.2$~mm), using the trap potential given by Eq.~\ref{exactpot}. The displacement in case of the HT is $z_0=0.1$~mm (black), $z_0=1$~mm (blue) and $z_0=10$~mm (green). (a)-(d) Trapping potentials along the $z$-axis (vertical direction) and $y$-axis (axial direction), comparing ODT power $P=3$~W and $P=0.3$~W. (e) Peak densities $n_0/N$ as function of $P$ and $T$, assuming $k_B T=(1/10)U_0$ (and therefore $T\propto P$).}
\label{HeHTpotentials}       
\end{figure}

The combined potential of a QMT and single beam ODT is given by:
\begin{equation}\label{exactpot}
U(x,y,z)=\mu B' \sqrt{x^2+4y^2+(z-z_0)^2}-\frac{2PC}{\pi w(y)^2}\exp\left[-2\frac{x^2+z^2}{w(y)^2}\right]+m g z, 
\end{equation}
where the first term is the QMT potential, the second term is the ODT potential, and the third term is the gravitational potential. In our case, the symmetry (strong) axes of the QMT and the ODT beam are along the $y$-axis. Here $B'$ is the magnetic field gradient along the weak axis of the QMT, $z_0$ is the vertical displacement of the QMT with respect to the ODT ($z_0>0$ means that the ODT is placed below the QMT center), $P$ is the power of the ODT beam, $C=\alpha_{\rm pol}/2\epsilon_0 c$ is a constant proportional to the polarizability $\alpha_{\rm pol}$ depending on the atomic species and used wavelength $\lambda$, $w(y)=w_{0}\sqrt{1+y^2/y_{R}^2}$, where $w_{0}$ and $y_{R}=\pi w_{0}^2/\lambda$ are the beam waist ($1/e^2$ radius) and the Rayleigh length, respectively. We use $\lambda=1557$~nm at which $C=1.88\times10^{-36}$ J/(W m$^{-2}$) \cite{notermans2014mwf}. 

For HT the magnetic force in the vertical direction should be equal or smaller than gravity, which means in our geometry that $B'\leq B'_{\rm lev}$. For $B' = B'_{\rm lev}$ the trap depth is always given by $U_0=2PC/(\pi w_0^2)$, whereas for $B' < B'_{\rm lev}$ (or $B'=0$ for a pure ODT) gravity leads to a reduction of the trap depth, which also depends on $P$, but this starts to be significant only for $P<100$~mW. The radial confinement is dominated by the ODT potential. For \He~the levitation gradient $B'_{\rm lev}=0.35$~G/cm is very small, due to the small mass and relatively large magnetic moment $\mu=2\mu_B$, where $\mu_B$ is the Bohr magneton. For comparison, the values for $^{87}$Rb are 15 or 30~G/cm, depending on the Zeeman state. Therefore the additional axial confinement provided by the QMT, compared to the pure ODT, is limited in the HT, and only for low ODT power $P$ the axial confinement is dominated by the QMT. In Fig.~\ref{HeHTpotentials}(a)-(d) we show the trapping potentials in both the radial and axial direction, for two different ODT powers, and $z_0$=0.1, 1 and 10~mm. Clearly, for $P=3$~W the axial confinement of the ODT is still significant, while for $P=0.3$~W the axial trapping potential is dominated by the QMT for sufficiently small $z_0$. 

The peak density $n_0$ of a thermal sample is obtained from numerically solving the integral $\int{\exp\left[-U(\vec{r})/k_B T\right]{\rm d}\vec{r}}=N/n_0$, where $U(\vec{r})$ is the trapping potential (Eq.~\ref{exactpot}). A comparison between the HT (for $z_0$=0.1, 1 and 10~mm) and ODT is given in Fig.~\ref{HeHTpotentials}(e), showing $n_0/N$ as function of the ODT power $P$ (and $T$), assuming that $T$ is determined by the trap depth according to $U_0=\eta k_B T$ (and therefore $T\propto P$), taking a typical value of $\eta=10$. Indeed, only for low $P$ the peak density $n_0$ in the HT is significantly higher than that of the ODT. Also the dependence on $z_0$ becomes more prominent for low $P$. For heavier atoms, like $^{87}$Rb, this regime is already reached at ODT powers well above 1~W. 

The small $B'_{\rm lev}$ limits the confinement also in an indirect way. Any stray magnetic field $B_{\rm offset}$ will shift the center of the QMT, and therefore affects the displacement with respect to the ODT center, such that $z_0'=z_0+B_{\rm offset}^{z}/B'$. Those stray magnetic fields can be compensated by additional bias fields, however, magnetic field fluctuations translate in a jitter of the QMT center, which limits the smallest displacement that can be chosen for which the atoms do not experience a magnetic field zero. At $B'=B'_{\rm lev}$ magnetic field fluctuations on the order of 10~mG will already give a jitter of 0.3~mm in the location of the QMT. Furthermore, even in the absence of magnetic field fluctuations, the magnetic field offset at the location of the potential minimum of the HT, $B_0=B' z_0$, has to be sufficiently large to provide a well-defined quantization axis, in order to suppress Majorana spin-flips and Penning ionization. For the heavier atoms this problem is much less severe. For instance, for $^{87}$Rb in the $F=2$, $m_F=2$ state a small displacement of about a waist, say $z_0=50$~$\mu$m, already gives an offset field of 75~mG, and a magnetic field fluctuation of 10~mG in the vertical direction only gives a jitter of 6~$\mu$m in the displacement. 

\section{Experimental setup}\label{expsetup}

\begin{figure}
\center
\resizebox{0.7\textwidth}{!}{%
  \includegraphics{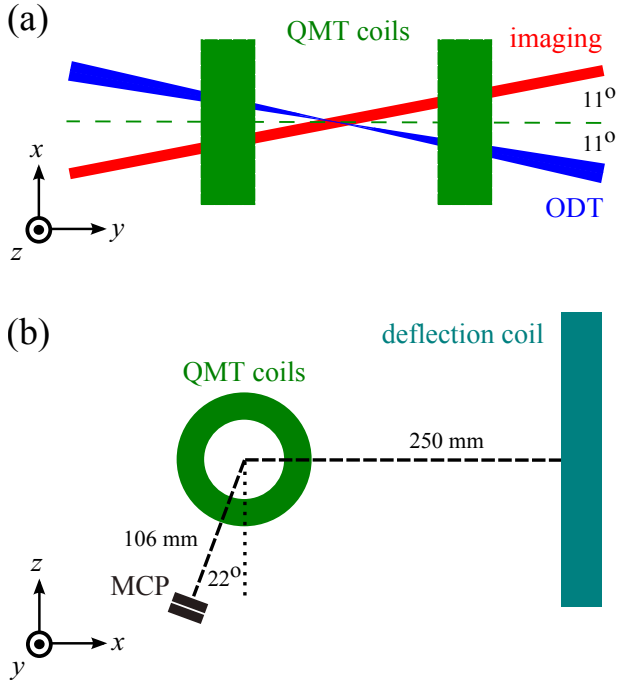}
}
\caption{(Color online) Schematics of our experimental setup, (a) showing the angles between the QMT axis, ODT beam and absorption imaging beam in the $x-y$ (horizontal) plane and (b) the location of the MCP detector and deflection coil in the $x-z$ plane.}
\label{setup}       
\end{figure}

The main part of our experimental setup has been described in Ref.~\cite{knoop2014umo}, while our HT (applied to $^{87}$Rb) has been discussed in Ref.~\cite{mishra2015epo}. Here we will summarize the most important features and mention changes made compared to previous work. A schematic of the HT is given in Fig.~\ref{setup}, showing the QMT coils, ODT beam, absorption imaging beam and micro-channel plate (MCP) detector. The axial direction of the QMT, the ODT beam and the absorption imaging beam are in the horizontal ($x-y$) plane. Absorption imaging is used to obtain information about the atom number and temperature, as well as the position of the QMT and ODT. We use an InGaAs camera (Xenics) with 30~$\mu$m pixel size and our imaging setup has a magnification of 0.5. 

We also record time-of-flight (TOF) spectra using a MCP detector, which is placed at a distance of 106~mm from the trap center, under an angle of 22$^\circ$ with respect to the direction of gravity, and has a diameter of 15~mm. For temperatures below 10~$\mu$K the ballistically expanding cloud, after switching off the trap, would not hit the MCP detector during its free fall. Therefore a short (10~ms) magnetic field gradient pulse is applied using a single ``deflection'' coil (see Fig.~\ref{setup}) to direct the atoms towards the MCP after release from the trap. The TOF spectra have a better resolution regarding temperature and observing BEC is much easier compared to expansion measurements with our absorption imaging system. However, the deflection field affects the TOF distribution and we need to use absorption imaging to calibrate the MCP detection regarding atom number and temperature.

Our single beam ODT has a waist $w_0=39.8\pm0.3$~$\mu$m (corresponding to a Rayleigh length of $y_R=3.2$~mm) and the maximum power available at the setup is about $P=3$~W, resulting in a maximum trap depth of $U_0=k_B\times160$~$\mu$K. We control the ODT power by an acousto-optical modulator (AOM), for which the output is coupled into a single mode fiber and sent to the experimental setup. The ODT beam enters the setup under an angle of 11$^\circ$ with respect to the QMT axis (see Fig.~\ref{setup}(a)), which leads to a reduction of the axial magnetic field gradient by a factor of $1-\sin(11^\circ)/2\approx 0.90$, but does not affect the vertical magnetic field gradient. 

\begin{figure}
\center
\resizebox{0.9\textwidth}{!}{%
  \includegraphics{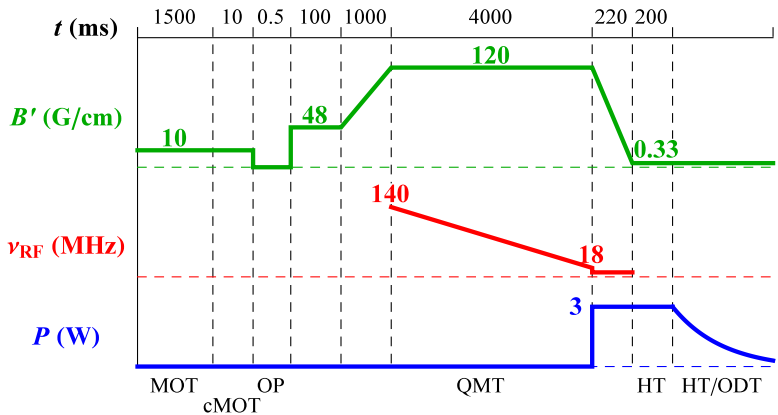}
}
\caption{(Color online) Overview of our experimental scheme for loading of and evaporation in the hybrid trap, showing the QMT gradient $B'$, radiofrequency $\nu_{\rm RF}$ and ODT power $P$.}
\label{expscheme}       
\end{figure}

An overview of our experimental scheme, in particular the magnetic field gradient $B'$, RF-frequency $\nu_{\rm RF}$ and ODT power $P$, is shown in Fig.~\ref{expscheme}. We use a liquid-nitrogen cooled dc-discharge source to produce a \He~beam, which is collimated and subsequently decelerated in a 2.5~m long Zeeman slower and loaded into a magneto-optical trap (MOT). An \textit{in vacuo} shutter after the collimation section is opened only during the loading time of the MOT. The pressure in the main vacuum chamber is $1\times10^{-10}$~mbar. We load about $5\times10^8$ \He~atoms at a temperature of about 1~mK within 1.5~s in the MOT, which consists of three retroreflected 2-inch laser beams at 1083~nm with a total power of $\sim$~45~mW and a large detuning of $-33$~MHz (21 linewidths), and a magnetic field gradient (weak axis) of 10~G/cm. Compared to Ref.~\cite{knoop2014umo} we have improved our atom number by increasing the MOT beam diameter from 1- to 2-inch. Afterwards we compress the cloud (the ``cMOT'' stage) by ramping down the detuning to $-6$~MHz in 10~ms, during which the power is reduced by a factor of ten, while keeping the same magnetic field gradient. After this cMOT stage we end up with $3\times 10^8$ atoms at a temperature of 260(10)~$\mu$K. Before loading in the QMT we optically pump (OP) the atoms into the $m$=+1 magnetic trappable state in 0.5~ms, during which the magnetic field gradient is switched off. Then we switch on abruptly the QMT at $B'=48$~G/cm and stay for 100~ms, and then ramp in 1~s to $B'$=120~G/cm. At this point we have about $1\times 10^8$ atoms at a temperature of about 1~mK. 

\begin{figure}
\center
\resizebox{0.8\textwidth}{!}{%
	\includegraphics{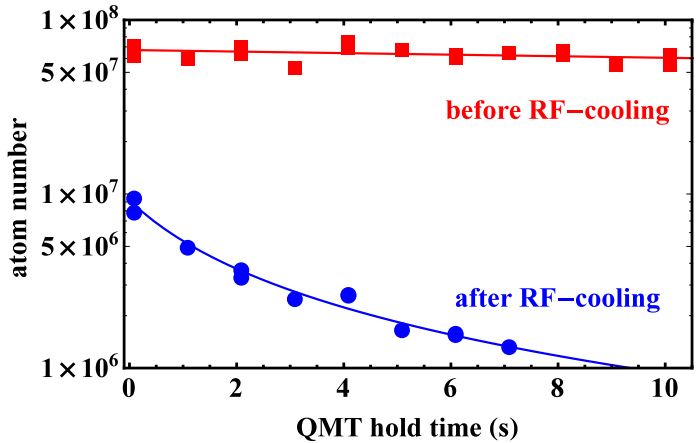}
}
\caption{(Color online) Lifetime in the QMT, before RF-cooling (red squares) and after RF-cooling (blue circles). Before RF-cooling we observe a slow, exponential decay of the number of trapped atoms, where the lifetime is limited by background collisions. After RF-cooling we observe a fast, non-exponential decay. Lines are the result of a loss model that includes Majorana loss and heating, fitted to the data.}
\label{QMTlifetimes}       
\end{figure}

After loading in the QMT, we apply RF-induced forced evaporative cooling, resulting in an effective trap depth $U_0^{\rm eff}=h\nu_{\rm RF}$. We generate the RF frequency $\nu_{\rm RF}$ by frequency doubling the output of a tunable 80~MHz function generator. After several amplification stages we send up to 5~W of RF power to a coil that is placed inside the vacuum chamber. We ramp down $\nu_{\rm RF}$ from 140~MHz to 18~MHz in 4~s, corresponding to a final trap depth of $k_B\times0.9$~mK. At this point the lifetime of the trapped atoms is only a few seconds, caused by Majorana spin-flips, which has a loss rate of $\Gamma_{\rm M}=\chi(\hbar/m)(2\mu B'/k_B T)^2$ \cite{dubessy2012rbe}. In Fig.~\ref{QMTlifetimes} we show the lifetime of the trapped atoms in the QMT, before and after RF cooling, where we have fitted the data with a loss model that includes Majorana loss and heating \cite{dubessy2012rbe,bouton2015fpo}. The lifetime data after RF-cooling are consistent with a temperature of $150-220$~$\mu$K for a $\chi$ factor between 0.1 and 0.2 \cite{dubessy2012rbe,bouton2015fpo}. To load the HT we switch on the ODT light at maximum power and ramp down the QMT gradient from $B'=120$~G/cm to $B'=0.33$~G/cm (just below $B'_{\rm lev}$) in 220~ms. During the ramp down $\nu_{\rm RF}$ is set at 9~MHz. Afterwards, an additional 200~ms is used to fine-adjust the bias fields to control the displacement (see Sect.~\ref{controldisplacement}). For loading of a pure ODT we switch off the QMT gradient during this last stage, while switching on a bias magnetic field in the axial direction to provide a quantization axis in order to keep the atoms spin-polarized.

\section{Results}\label{results}

\begin{figure}
\center
\resizebox{0.7\textwidth}{!}{%
	\includegraphics{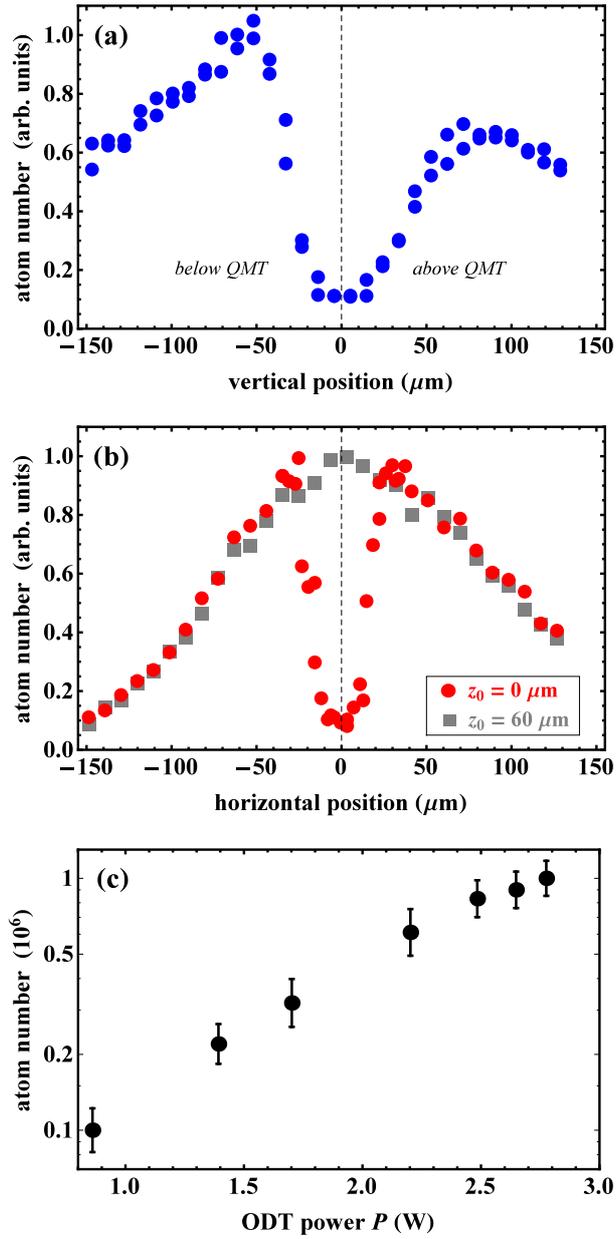}
	}
\caption{(Color online) Number of atoms loaded in the hybrid trap at $P=2.8$~W as function of position of the ODT (using a piezo-mirror) with respect to the QMT center in the vertical direction (a) and horizontal direction (b). The gray data in (b) shows a scan in the horizontal, while the vertical position is 60~$\mu$m below the QMT center. Panel (c): number of atoms loaded in the hybrid trap as function of ODT power $P$, with vertical displacement of 60~$\mu$m.}
\label{piezoscan}       
\end{figure}

\subsection{Loading of HT or ODT}

For the alignment of the ODT beam we use a piezo-mirror to scan in both the horizontal ($x$-axis) and vertical ($z$-axis) direction and monitor the number of atoms loaded into the HT or ODT. Typical measurements are shown in Fig.~\ref{piezoscan}(a)-(b). In both directions one finds a minimum when the ODT is located at the center of the QMT, which is due to Majorana loss, and the width of the loss feature is on the order of the waist. We position our ODT about 60~$\mu$m below the QMT center. We obtain up to $1\times10^6$ atoms at a temperature of 14~$\mu$K for both HT and ODT. The transfer efficiency from the QMT to the HT or ODT is about $5-10$\%.

We have measured the number of loaded atoms for different initial ODT powers in order to investigate to what extent we are limited by our maximum ODT power of 2.8~W. The results are shown in Fig.~\ref{piezoscan}(c). The number of loaded atoms does not fully saturate, meaning that the number of transferred atoms is limited by our ODT power. At the maximum power we measure a 1/e trapping lifetime of more than 20~s of the HT and pure ODT, at which the calculated off-resonant photon scattering at 1557~nm is (6~s)$^{-1}$, but the recoil temperature of 2~$\mu$K is much smaller than the trap depth. 

\subsection{Control of QMT displacement in HT}\label{controldisplacement}

\begin{figure}
\center
\resizebox{0.8\textwidth}{!}{%
	\includegraphics{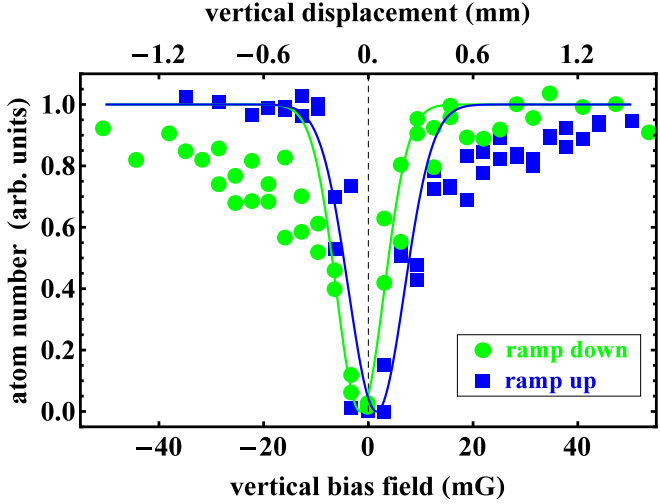}
	}
\caption{(Color online) Number of atoms in the hybrid trap as function of an applied vertical bias field that controls the positions of the QMT center with respect to the ODT. After the vertical bias field is set to its final value the total hold time is 300~ms, during which the ODT power is ramped down from $P=2.8$~W to 2.2~W. The loss resonance occurs due to Majorana spin-flips when the center of the QMT is overlapped with the ODT. An asymmetry and broadening of the loss feature is caused by the slow response of the magnetic field coil. Two sets of data are shown, for which we ramp up from a value just below the resonance (blue squares) or ramp down from above the resonance (green circles). The solid lines are Gaussian fits taking into account only the fast slopes. The resonance position is determined by the average of the two Gaussian centers. With the magnetic field gradient of 0.33~G/cm in the vertical direction, the vertical bias field can be converted into vertical displacement using $z_0=B_z/B'$.}
\label{verticalbiasscan}       
\end{figure}

During the QMT stage magnetic bias fields are set to minimize stray magnetic fields at the center of the trap and we use the piezo-mirror to adjust the displacement of the ODT with respect to the QMT center. However, in the HT this displacement becomes very sensitive to bias fields, which provides a much broader tuning range (up to a few cm) compared to the piezo-mirror (limited to 0.3~mm). Therefore, once the atoms are loaded in the HT, we fine-adjust these magnetic bias fields to set the displacement of the QMT center with respect to the ODT, while keeping the piezo-mirror at the optimal loading condition (see Fig.~\ref{piezoscan}(a)-(b)). In scanning the bias fields in the $x$- and $z$-direction we do observe loss resonances at which atoms are lost on the time scale of 100~ms. Again, those losses are due to Majorana spin-flips in which the atoms can leave the HT in the axial direction. In Fig.~\ref{verticalbiasscan} we show the number of atoms in the HT as function of vertical bias field after a hold time of 300~ms, in which we jump to the field value from below and above the resonance position. The asymmetry of the loss resonances is due to the slow response of the magnetic field coils. However, by fitting Gaussian distributions to the two scans, in which the slow rising part of the data is omitted, the centers and actual widths can be determined, where the actual center is taken as the average of the two centers. We find a $1/e^2$ width of 5~mG and the center is reproducible within 3~mG on a day-to-day basis. 

While the center of the loss resonance fixes $B_z=0$, the displacement is simply given by $z_0=B_z/B'$. Here $B'=0.33$~G/cm, which means that the width of 5~mG already corresponds to 0.15~mm. This width is much larger than observed in scanning the ODT by means of the piezo-mirror to find the optimum loading conditions (see Fig.~\ref{piezoscan}), where the width is on the order of the waist. This can be explained by assuming that the atoms are transferred from the QMT to the HT already at $B'>>B'_{\rm lev}$, where the sensitivity to magnetic bias fields is much less. It also means that the displacement assigned in the measurement of Fig.~\ref{piezoscan} only holds for the loading, not the final HT. While the piezo-mirror is optimized on the loading, we use the magnetic bias fields to set the displacement of the final HT, which also covers a much broader range. In the $x$-direction we set the bias field ``on resonance'' ($x_0=0$), and control the displacement with the vertical bias field. In order not be effected by Majorana spin-flips a displacement larger than 0.3~mm has to be chosen, which limits the axial confinement and the peak density (see Fig.~~\ref{HeHTpotentials}), and the displacement jitters by 0.15~mm. Again, such constraints are essentially not present for the heavier atoms, in which the displacement can be chosen to be on the order of a waist and the jitter is much smaller than the waist.

\subsection{Evaporative cooling in HT and ODT}\label{evaporativecooling}

\begin{figure}
\center
\resizebox{0.8\textwidth}{!}{%
	\includegraphics{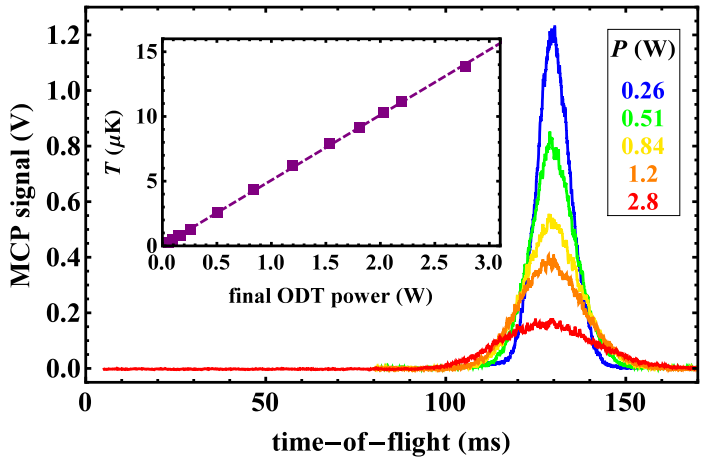}
	}
\caption{(Color online) Forced evaporative cooling in the pure ODT, showing TOF spectra obtained by the MCP detector for different final ODT powers. The inset shows the temperature, obtained from the width of the TOF spectra, as function of final ODT powers (dashed line is a linear fit), from which a truncation parameter of $\eta=U_0/k_B T=11$ is obtained. Each TOF spectra is an average over 5 experimental runs.}
\label{MCPTOFoverview}       
\end{figure}

After loading the HT or ODT we perform forced evaporative cooling to lower the temperature and increase the phase-space density by ramping down the ODT power. A sample of TOF spectra obtained by the MCP detector is shown in Fig.~\ref{MCPTOFoverview}. In the regime where the kinetic energy is much smaller than the gravitational energy, i.\,e.\, $k_B T << m g h$, where $h$ is the height difference between the trap and the detector, the TOF spectrum of a thermal sample is described by a Gaussian distribution with a width that is proportional to the square-root of the temperature. Indeed, for lower ODT power we observe a narrowing of the distribution, but also an increase of the signal, which is due to the finite size of the MCP detector. For a thermalized sample the temperature is proportional to the trap depth $U_0=2PC/(\pi w_0^2)$ via the truncation parameter $\eta=U_0/k_B T$. Indeed, we find the width to be proportional to the square-root of the ODT power $P$. By fitting the relation $\sigma=a\sqrt{P}+b$ to the data we find a small offset $b$ of about 1~ms, which is probably caused by the magnetic field gradient pulse that we apply after release from the trap to direct the atoms towards the MCP. Before converting the width into temperature we correct for this offset. Absolute calibration of the temperature is done by absorption imaging at $P=2.8$~W, at which we have measured $13.9\pm 0.2$~$\mu$K. The result is shown in the inset of Fig.~\ref{MCPTOFoverview}, in which the data shows a linear behavior for the full range of ODT powers. From the slope one directly obtains $\eta$, which turns out to be 11 for both HT and ODT.

\begin{figure}
\center
\resizebox{0.8\textwidth}{!}{%
	\includegraphics{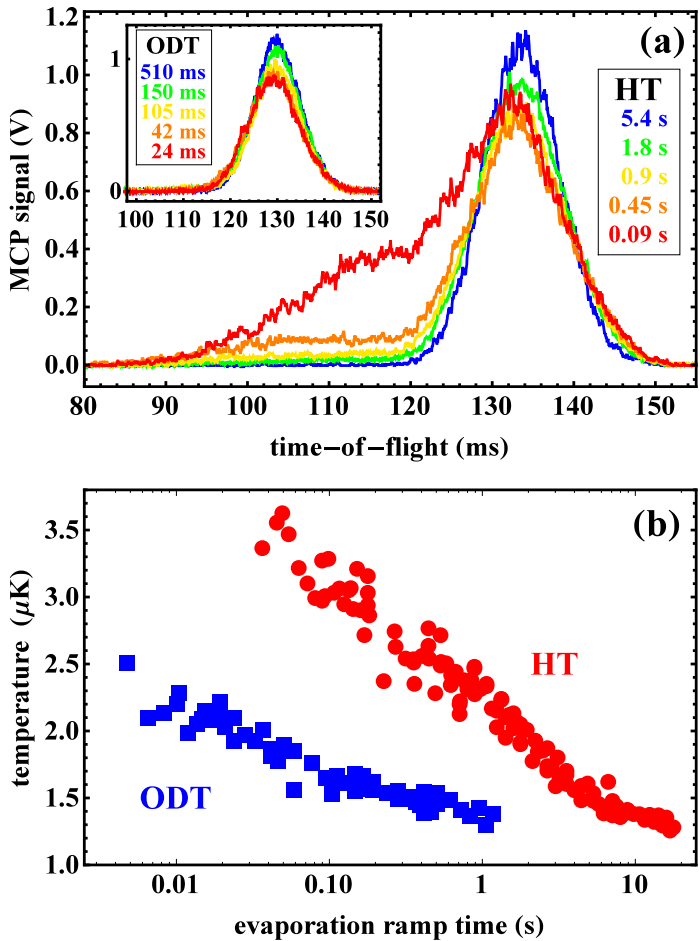}
	}
\caption{(Color online) Forced evaporative cooling in HT (with $z_0=1.5$~mm) and ODT, ramping down the ODT power from $P=2.8$~W to 0.26~W. (a) TOF spectra for HT and ODT (inset) for different evaporation ramp times (each TOF spectrum is an average over 2 to 4 experimental runs). (b) Temperature as function of ramp time for HT and ODT. The initial temperature at $P=2.8$~W is 14~$\mu$K.}
\label{ramptimeHTODT}       
\end{figure}

First we investigate the time-scale of forced evaporative cooling, comparing HT ($z_0=1.5$~mm) and ODT. We note that the initial conditions are the same in terms of temperature (14~$\mu$K) and atom number ($1\times 10^6$), while the peak density and the collision rate are slightly higher for the HT ($2.1\times 10^{12}$~cm$^{-3}$ and 410 s$^{-1}$ for HT; $1.7\times 10^{12}$~cm$^{-3}$ and 320 s$^{-1}$ for ODT). We ramp down the ODT power from $P=2.8$~W to 0.26~W for variable ramp times. A sample of TOF spectra are shown in Fig.~\ref{ramptimeHTODT}(a), for the HT and ODT (inset). Two striking observations can be made: first of all the time-scale for thermalization is much shorter for the ODT than the HT, and secondly, a shoulder on the left side of the TOF peak appears for short evaporation ramp times in the HT, corresponding to ``hot atoms'' that remain trapped. In Fig.~\ref{ramptimeHTODT}(a) we show the temperature, obtained by a fit of the main TOF peak, as function of evaporation ramp time. The initial temperature at $P=2.8$~W is 14~$\mu$K, while the final temperature at $P=0.26$~W is 1.3~$\mu$K. For the ODT a temperature of 1.5~$\mu$K is reached in 100~ms, which for the HT it takes about 3~s. We have observed similar behavior for a HT at a much larger displacement of $z_0=15$~mm, and a HT at about half the levitation gradient ($B'=0.16$~G/cm, $z_0=1.2$~mm).

We explain these observation by the dimensionality of evaporation: while in the ODT the atoms can escape in all directions, in the HT they can only escape downwards. Therefore the removal of ``hot atoms'' takes much longer, even though the collision rate in the HT at $P=0.26$~W is about twice that of the ODT ($1.5\times 10^{12}$~cm$^{-3}$ compared to $7\times 10^{11}$~cm$^{-3}$). The required evaporation ramp time for the ODT compares quite well with a calculated thermalization rate at $P=0.26$~W of (70~ms)$^{-1}$, while the axial trap frequency is $2\pi \times 12$~Hz. The longest time scale of the HT is the axial trap frequency of $2\pi \times 27$~Hz.

\begin{figure}
\center
\resizebox{0.8\textwidth}{!}{%
	\includegraphics{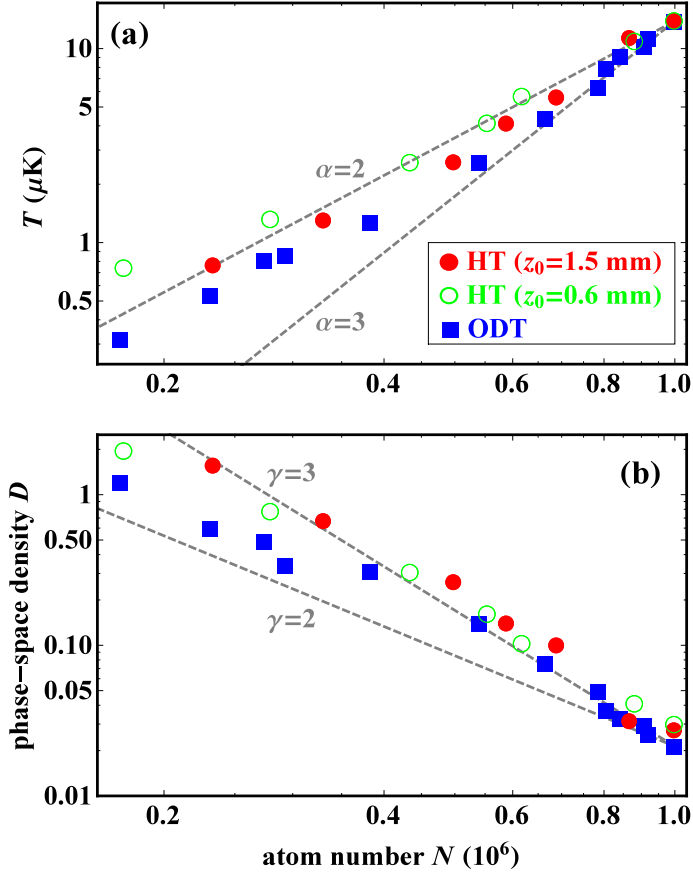}
	}
\caption{(Color online) Evaporative cooling in the HT $z_0=0.6$~mm and $z_0=1.5$~mm and ODT, showing (a) the temperature and (b) phase-space density as function the atom number. The gray dashed lines indicate the efficiency in terms of $\alpha={\rm d}[\log T]/{\rm d}[\log N]$ and $\gamma=-{\rm d}[\log D]/{\rm d}[\log N]$ parameters.}
\label{evaporativecoolingHTODT}       
\end{figure}

After having determined the appropriate ramp times (also for lower final ODT powers) we study the efficiency of evaporation by measuring the atom number and temperature for different final ODT power. In Fig.~\ref{evaporativecoolingHTODT}(a) we show the temperature as function of atom number in the HT for two displacements, $z_0=0.6$~mm and $z_0=1.5$~mm, and the ODT. Here the efficiency is typically quantified as $\alpha={\rm d}[\log T]/{\rm d}[\log N]$, and we observe a $\alpha$ parameter between 2 and 3. For the same final temperature we obtain the highest atom number for ODT and the lowest one for the HT with the smallest displacement. 

In Fig.~\ref{evaporativecoolingHTODT}(b) we present the phase-space density $D=n_0\lambda_{\rm dB}^3$ (where $\lambda_{\rm dB}=h/\sqrt{2\pi m k_B T}$ is the de Broglie wavelength) as function of atom number, for which the peak density $n_0$ is calculated using numerical integration of Eq.~\ref{exactpot} (see Sect.~\ref{HT}). Here the efficiency is typically quantified as $\gamma=-{\rm d}[\log D]/{\rm d}[\log N]$, and we observe a $\gamma$ parameter between 2 and 3. We reach the onset of BEC ($D>1$) in all three cases, with the most atoms of $2\times 10^5$ for the HT with $z_0=1.5$~mm, showing a slightly better performance than the one with $z_0=0.6$~mm. But even for the single beam ODT we obtain efficient evaporation up to $D=1$ with $\gamma>2$, which is probably explained by the sufficiently large axial trap frequencies of at least a few Hz.

\begin{figure}
\center
\resizebox{0.8\textwidth}{!}{%
	\includegraphics{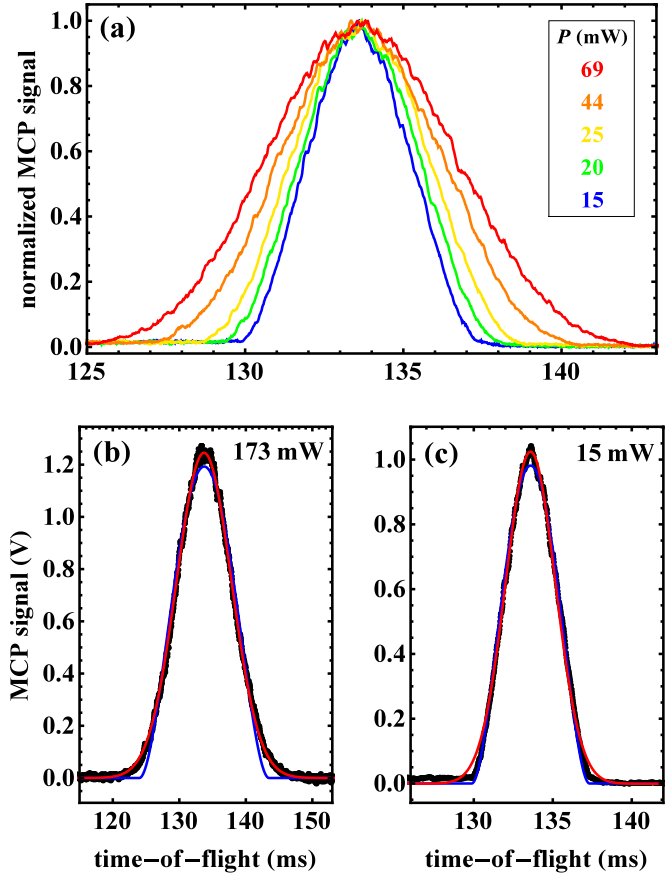}
	}
\caption{(Color online) (a) Normalized TOF spectra of the HT with $z_0=1.5$~mm for different final ODT powers at which $D>1$, showing a sharpening of the wings of the distributions. Panel (b) and (c) show TOF spectra for $P=173$~mW and $P=15$~mW, for which $D\approx1$ and $D>>1$, respectively, together with Gaussian fits (red) and BEC fits (blue, Eq.~\ref{BECformula}). Each TOF spectrum is an average over 5 experimental runs. The shot-to-shot fluctuation in the signal height is less than 5~\%.}
\label{HTBECTOF}       
\end{figure}

The appearance of BEC can be observed in the TOF spectra obtained by MCP detection, as it provides a distinct TOF distribution described by an inverted parabola,
\begin{equation}\label{BECformula}
\Phi_{\rm BEC}(t)\propto\max\left[0,1-\left(\frac{t-t_0}{\sigma}\right)^2\right]^2,
\end{equation}
where the width $\sigma$ is given by the chemical potential \cite{robert2001bec,tychkov2006mtb,dall2007bec}. Compared to the Gaussian distribution associated with a thermal sample, a BEC gives rise to much sharper wings of the distribution. In our case, the magnetic field gradient pulse, which is required to direct the atoms towards the MCP after release from the trap, affects the narrow BEC TOF peak, and we do not obtain clear bimodal distributions in which the thermal and BEC part can easily be distinguished. Nonetheless, for the lowest final ODT powers we do observe a sharpening of the wings of the TOF distribution. In Fig.~\ref{HTBECTOF}(a) we show a series of TOF spectra of the HT with $z_0=1.5$~mm for decreasing ODT power, for which $D>1$. The narrowing of the TOF distributions can be explained by both a decreasing temperature and a decreasing chemical potential, due to reduction of trap frequencies. In Fig.~\ref{HTBECTOF}(b)-(c) we show pure thermal and BEC fits to the situation of (b) $D\approx1$ (c) and $D>>1$. Indeed, the BEC distribution describes the TOF spectrum of panel (c) quite well, much better than the thermal fit. From this observation we estimate that we have reached a pure BEC of $5\times10^4$ atoms. 

\section{Conclusions}\label{conclusions}

We have achieved Bose-Einstein condensation (BEC) of metastable triplet helium atoms via RF-induced evaporative cooling in a quadrupole magnetic trap, transfered to a single beam hybrid trap, and subsequent evaporative cooling in both the hybrid trap and pure optical dipole trap, using only moderate ODT power of less than 3~W. While evaporation in the pure ODT is much faster than that in the HT, a larger BEC is obtained in the HT. We reach the onset of BEC with $2\times 10^5$ atoms and a pure BEC of $5\times 10^4$ atoms. We observe small shot-to-shot fluctuations (less than 5\%) and excellent day-to-day reproducibility. The total experimental cycle duration is between 8 to 10~s. This all could be improved by the implementation of a bright molasses \cite{chang2014tdl} and gray molasses \cite{bouton2015fpo} before loading the QMT, which would provide better initial conditions of RF-induced forced evaporative cooling in the QMT and therefore shorten the duration of that stage, and finally lead to a larger BEC. Our present work provides the most simple scheme so far to obtain a \He~BEC, using limited experimental infrastructure. It also shows that a single beam hybrid trap can be applied for a light atom such as helium, despite several challenges caused by its small mass and small levitation gradient. 

\begin{acknowledgement}
We acknowledge Rob Kortekaas for technical support. This work was financially supported by the Netherlands Organization for Scientific Research (NWO) via a VIDI grant (680-47-511) and the Dutch Foundation for Fundamental Research on Matter (FOM) via a Projectruimte grant (11PR2905). 
\end{acknowledgement}


\begin{thebibliography}{10}

\bibitem{lin2009rpo}
Y.-J. Lin, A.~R. Perry, R.~L. Compton, I.~B. Spielman, and J.~V. Porto,
\newblock Phys.\ Rev.\ A {\bf 79}, 063631 (2009).

\bibitem{kleinebuning2010asg}
G.~{Kleine B\"{u}ning}, J.Will, W.~Ertmer, C.~Klempt, and J.~Arlt,
\newblock Appl.\ Phys.\ B {\bf 100}, 117 (2010).

\bibitem{mishra2015epo}
H.~P. Mishra, A.~S. Flores, W.~Vassen, and S.~Knoop,
\newblock Eur.\ Phys.\ J.\ D {\bf 69}, 52 (2015).

\bibitem{petrich1995stc}
W.~Petrich, M.~H. Anderson, J.~R. Ensher, and E.~A. Cornell,
\newblock Phys.\ Rev.\ Lett. {\bf 74}, 3352 (1995).

\bibitem{dubessy2012rbe}
R.~Dubessy, K.~Merloti, L.~Longchambon, P.-E. Pottie, T.~Liennard, A.~Perrin,
  V.~Lorent, and H.~Perrin,
\newblock Phys.\ Rev.\ A {\bf 85}, 013643 (2012).

\bibitem{schellekens2005hbt}
M.~Schellekens, R.~Hoppeler, A.~Perrin, J.~{Viana Gomes}, D.~Boiron, A.~Aspect,
  and C.~I. Westbrook,
\newblock Science {\bf 310}, 648 (2005).

\bibitem{jeltes2007cot}
T.~Jeltes, J.~M. McNamara, W.~Hogervorst, W.~Vassen, V.~Krachmalnicoff,
  M.~Schellekens, A.~Perrin, H.~Chang, D.~Boiron, A.~Aspect, and C.~I.
  Westbrook,
\newblock Nature {\bf 445}, 402 (2007).

\bibitem{hodgman2011dmo}
S.~S. Hodgman, R.~G. Dall, A.~G. Manning, K.~G.~H. Baldwin, and A.~G. Truscott,
\newblock Science {\bf 331}, 1046 (2011).

\bibitem{lopes2015aho}
R.~Lopes, A.~Imanaliev, A.~Aspect, M.~Cheneau, D.~Boiron, and C.~I. Westbrook,
\newblock Nature {\bf 520}, 66 (2015).

\bibitem{manning2015wdc}
A.~G. Manning, R.~I. Khakimov, R.~G. Dall, and A.~G. Truscott,
\newblock Nature Phys. {\bf 11}, 539 (2015).

\bibitem{rooij2011fmi}
R.~{van Rooij}, J.~S. Borbely, J.~Simonet, M.~D. Hoogerland, K.~S.~E. Eikema,
  R.~A. Rozendaal, and W.~Vassen,
\newblock Science {\bf 333}, 196 (2011).

\bibitem{notermans2014hps}
R.~P. M. J.~W. Notermans and W.~Vassen,
\newblock Phys.\ Rev.\ Lett. {\bf 112}, 253002 (2014).

\bibitem{henson2015pmo}
B.~M. Henson, R.~I. Khakimov, R.~G. Dall, K.~G.~H. Baldwin, L.-Y. Tang, and
  A.~G. Truscott,
\newblock Phys.\ Rev.\ Lett. {\bf 115}, 043004 (2015).

\bibitem{robert2001bec}
A.~Robert, O.~Sirjean, A.~Browaeys, J.~Poupard, S.~Nowak, D.~Boiron, C.~I.
  Westbrook, and A.~Aspect,
\newblock Science {\bf 292}, 461 (2001).

\bibitem{pereira2001bec}
F.~{Pereira Dos Santos}, J.~L\'eonard, J.~Wang, C.~J. Barrelet, F.~Perales,
  E.~Rasel, C.~S. Unnikrishnan, M.~Leduc, and C.~Cohen-Tannoudji,
\newblock Phys.\ Rev.\ Lett. {\bf 86}, 3459 (2001).

\bibitem{pereira2002poa}
F.~{Pereira Dos Santos}, J.~L\'eonard, J.~Wang, C.~J. Barrelet, F.~Perales,
  E.~Rasel, C.~S. Unnikrishnan, M.~Leduc, and C.~Cohen-Tannoudji,
\newblock Eur.\ Phys.\ J.\ D {\bf 19}, 103 (2002).

\bibitem{tychkov2006mtb}
A.~S. Tychkov, T.~Jeltes, J.~M. McNamara, P.~J.~J. Tol, N.~Herschbach,
  W.~Hogervorst, and W.~Vassen,
\newblock Phys.\ Rev.\ A {\bf 73}, 031603(R) (2006).

\bibitem{dall2007bec}
R.~G. Dall and A.~G. Truscott,
\newblock Opt. Commun. {\bf 270}, 255 (2007).

\bibitem{doret2009bgc}
S.~C. Doret, C.~B. Connolly, W.~Ketterle, and J.~M. Doyle,
\newblock Phys.\ Rev.\ Lett. {\bf 103}, 103005 (2009).

\bibitem{keller2014bec}
M.~Keller, M.~Kotyrba, F.~Leupold, M.~Singh, M.~Ebner, and A.~Zeilinger,
\newblock Phys.\ Rev.\ A {\bf 90}, 063607 (2014).

\bibitem{partridge2010bec}
G.~B. Partridge, J.-C. Jaskula, M.~Bonneau, D.~Boiron, and C.~I. Westbrook,
\newblock Phys.\ Rev.\ A {\bf 81}, 053631 (2010).

\bibitem{dall2010tmi}
R.~G. Dall, S.~S. Hodgman, M.~T. Johnsson, K.~G.~H. Baldwin, and A.~G.
  Truscott,
\newblock Phys.\ Rev.\ A {\bf 81}, 011602(R) (2010).

\bibitem{vassen2012cat}
W.~Vassen, C.~Cohen-Tannoudji, M.~Leduc, D.~Boiron, C.~Westbrook, A.~Truscott,
  K.~Baldwin, G.~Birkl, P.~Cancio, and M.~Trippenbach,
\newblock Rev.\ Mod.\ Phys. {\bf 84}, 175 (2012).

\bibitem{bouton2015fpo}
Q.~Bouton, R.~Chang, A.~L. Hoendervanger, F.~Nogrette, A.~Aspect, C.~I.
  Westbrook, and D.~Cl\'{e}ment,
\newblock Phys.\ Rev.\ A {\bf 91}, 061402(R) (2015).

\bibitem{notermans2014mwf}
R.~P. M. J.~W. Notermans, R.~J. Rengelink, K.~A.~H. {van Leeuwen}, and
  W.~Vassen,
\newblock Phys.\ Rev.\ A {\bf 90}, 052508 (2014).

\bibitem{knoop2014umo}
S.~Knoop, P.~S. \.{Z}uchowski, D.~K\c{e}dziera, {\L}.~Mentel, M.~Puchalski,
  H.~P. Mishra, A.~S. Flores, and W.~Vassen,
\newblock Phys.\ Rev.\ A {\bf 90}, 022709 (2014).

\bibitem{chang2014tdl}
R.~Chang, A.~L. Hoendervanger, Q.~Bouton, Y.~Fang, T.~Klafka, K.~Audo,
  A.~Aspect, C.~I. Westbrook, and D.~Cl\'{e}ment,
\newblock Phys.\ Rev.\ A {\bf 90}, 063407 (2014).

\end{thebibliography}
\end{document}